# Visualizing delocalized correlated electronic states in twisted double bilayer graphene

Canxun Zhang,[1, 2, 3, 7] Tiancong Zhu,[1, 2, 7] Salman Kahn,[1, 2, 7] Shaowei Li,[1, 2, 3] Birui Yang,[1] Charlotte Herbig,[1] Xuehao Wu,[1] Hongyuan Li,[1, 2] Kenji Watanabe,[4] Takashi Taniguchi,[5] Stefano Cabrini,[6] Alex Zettl,[1, 2, 3] Michael P. Zaletel,[1, 2]* Feng Wang,[1, 2, 3]* and Michael F. Crommie[1, 2, 3]*

[1]Department of Physics, University of California, Berkeley, CA 94720, USA.

[2]Materials Sciences Division, Lawrence Berkeley Laboratory, Berkeley, CA 94720, USA.

[3]Kavli Energy NanoScience Institute, University of California, Berkeley, CA 94720, USA.

[4]Research Center for Functional Materials, National Institute for Materials Science, 1-1 Namiki, Tsukuba 305-0044, Japan.

[5]International Center for Materials Nanoarchitectonics, National Institute for Materials Science, 1-1 Namiki, Tsukuba 305-0044, Japan.

[6]Molecular Foundry, Lawrence Berkeley Laboratory, Berkeley, CA 94720, USA.

[7]These authors contributed equally.

*Correspondence to: mikezaletel@berkeley.edu (M.P.Z.); fengwang76@berkeley.edu (F.W.); crommie@berkeley.edu (M.F.C.)

**Abstract**

The discovery of interaction-driven insulating and superconducting phases in moiré van der Waals heterostructures has sparked considerable interest in understanding the novel correlated physics of these systems. While a significant number of studies have focused on twisted bilayer graphene, correlated insulating states and a superconductivity-like transition up to 12 K have been reported in recent transport measurements of twisted double bilayer graphene. Here we present a scanning tunneling microscopy and spectroscopy study of gate-tunable twisted double bilayer graphene devices. We observe splitting of the van Hove singularity peak by ~20 meV at half-filling of the conduction flat band, with a corresponding reduction of the local density of states at the Fermi level. By mapping the tunneling differential conductance we show that this correlated system exhibits energetically split states that are spatially delocalized throughout the different regions in the moiré unit cell, inconsistent with order originating solely from onsite Coulomb repulsion within strongly-localized orbitals. We have performed self-consistent Hartree-Fock calculations that suggest exchange-driven spontaneous symmetry breaking in the degenerate conduction flat band is the origin of the observed correlated state. Our results provide new insight into the nature of electron-electron interactions in twisted double bilayer graphene and related moiré systems.

**Introduction**

The stacking of atomically-thin van der Waals (vdW) materials provides an elegant platform for studying correlated electronic states. Moiré superlattices formed by lattice misalignment between adjacent vdW sheets can create narrow mini-bands with width comparable to or even smaller than the Coulomb interaction energy,[1,2] leading to the emergence of various correlated phases. Experimental signatures of interaction-driven electronic states in moiré vdW stacks were first observed in magic-angle twisted bilayer graphene (tBLG), where the coexistence of insulating and superconducting phases resembles the phase diagram of high temperature cuprate superconductors.[3-7] More exotic phases, such as orbital ferromagnets and correlated Chern insulators, were later reported in other moiré vdW systems.[8-15] Transport measurements on twisted double bilayer graphene (tDBLG), which consists of two sheets of Bernal-stacked bilayer graphene with a small rotational misalignment, have shown new correlated features such as enhancement of a correlation gap under external magnetic field and a superconductivity-like critical transition with onset temperature as high as 12 K.[16-20] The precise nature of the electronic states in tDBLG, however, remains elusive, and the lack of spatially-resolved electronic structure data creates challenges in modeling its rich correlation physics.

Here we use scanning tunneling microscopy and spectroscopy (STM/STS) to study the local electronic structure of tDBLG. Fabrication of gate-tunable tDBLG devices allows us to continuously change the carrier density in our samples and to probe their local density of states (LDOS) for different electron filling levels. We observe that the low-energy electronic structure of tDBLG is dominated by two narrow moiré mini-bands that we refer to as the conduction flat band (CFB) and the valence flat band (VFB), each of which accommodates four electrons per moiré unit cell due to spin and valley degeneracies. The energy separation of these two bands can be tuned by an applied vertical electric field. When the CFB is half-filled we observe an interaction-driven reduction of LDOS at the Fermi level, consistent with an emergent correlated insulating phase first detected in transport studies.[16-20] In contrast to the electronic structure of tBLG,[1,2,21-25] we find that flat band wavefunctions in tDBLG are delocalized in real space and that the correlation-induced LDOS reduction is present everywhere in the moiré unit cell. This spatially-extended correlated state suggests that a picture in which the insulating phase is caused by strong onsite-repulsion between highly-localized electron orbitals is inadequate to explain correlation effects in tDBLG. Hartree-Fock calculations based on a continuum model, on the other hand, exhibit good agreement with our experimental data and reveal a large non-local exchange interaction. This points to an exchange-driven spontaneous symmetry breaking mechanism involving the four-fold degenerate CFB as the origin of observed splitting in tDBLG, analogous to integer quantum Hall ferromagnetism.[26,27]

**Results**

Our tDBLG samples were fabricated using a tear-and-stack technique with a hexagonal boron nitride (hBN) substrate and deposited onto a $SiO_2/Si$ wafer (Methods). Figure 1a shows a sketch of the device scheme and Fig. 1b shows an optical microscope picture of a typical sample. The presence of hBN and $SiO_2$ dielectric layers allows us to apply a voltage $V_G$ to the Si back-gate to change the carrier density $n$ and the vertical electric field $E$ in the tDBLG stack. The tDBLG samples were annealed in ultra-high vacuum before being loaded into the STM system at $T = 4.7$ K for measurement (Methods). Figure 1c shows a representative STM topographic image (the inset shows a zoom-in image with graphene lattice). The moiré wavelength of ~13 nm corresponds to a local twist angle of $\theta \approx 1.08°$, but the moiré pattern is not perfectly three-fold symmetric due to external strain. Variation of the moiré wavelength in different directions allows us to estimate a strain of ~0.2 % in this sample. As seen in Fig. 1c we observe three regions that have different apparent heights within each moiré unit cell. We identify these as the three different possible stacking regions of tDBLG: ABBC, ABCA, and ABAB (sketched in Fig. 1d; see Supplementary Note 1).

We characterized the electronic structure of our tDBLG samples by performing d$I$/d$V$ spectroscopy. Before each set of measurements our STM tips were calibrated against the Cu(111) Shockley surface state to ensure that they were free of artifacts known to arise from loosely bound adsorbates and clusters. d$I$/d$V$ spectra for tDBLG devices measured over large bias ranges using calibrated tips always exhibited a strong enhancement of the tunneling signal for $|V_{Bias}| > 60$ mV (Supplementary Fig. 1), a familiar graphene effect known to arise due to phonon-mediated inelastic tunneling.[28,29] The signal from the elastic tunneling channel is typically weak compared to the inelastic signal. Here we focus primarily on the elastic signal within the bias range $-60$ mV $\leq V_{Bias} \leq 60$ mV, both because correlation effects are expected to be strongest near $V_{Bias} = 0$ mV (the Fermi level) and to avoid inelastic broadening effects.[30] Figure 2a shows typical d$I$/d$V$ spectra obtained in the ABAB region at different gate voltages. For $V_G = 0$ V we observe a prominent peak centered just above $V_{Bias} = 0$ V that exhibits a shoulder at lower energy. This feature can be fit by the sum of two Lorentzian peaks centered at $V_{Bias} = 8$ mV and –4 mV, both of which have a full width at half maximum (FWHM) of ~23 meV (dashed lines in Fig. 2a). Due to the energetic narrowness of these peaks we label them as the conduction flat band (CFB) and the valence flat band (VFB), respectively. Increasing the gate voltage causes the sample to become more electron-doped and induces the CFB peak to slowly shift downward in energy. The VFB peak, on the other hand, more rapidly moves toward lower energy and away from the CFB for higher gate voltage. For $V_G \geq 45$ V the VFB signal shifts out of our measurement bias range $|V_{Bias}| \leq 60$ mV and only the CFB peak can be observed.

The energetic separation between the CFB and VFB peaks at finite gate voltages allows us to measure the spatial profile of each flat band feature individually. Figure 3a, for example, compares point spectra obtained for the CFB state in the ABBC (blue), ABCA (purple), and ABAB (orange) stacking regions at a gate voltage of $V_G = 60$ V. The CFB peak at $V_{Bias} = -20$ mV appears in all three regions with nearly equal intensity. In the ABBC region another peak appears at a higher energy of $V_{Bias} = 6$ mV that has a larger width than the CFB peak. We refer to this feature as the "remote conduction band" (RCB). To better visualize the spatial distributions of the CFB and RCB wavefunctions, we obtained d$I$/d$V$ spectroscopy grids over an area containing several moiré unit cells. Figure 3b,c show d$I$/d$V$ maps obtained at the CFB peak energy ($V_{bias} = -20$ mV) and the RCB peak energy ($V_{bias} = 6$ mV) plotted using the same color scale. We observe that the CFB signal is highly delocalized with only slight amplitude variation within the moiré unit cell (Fig. 3b). The RCB signal, on the other hand, is more strongly modulated and has its highest amplitude in the ABBC region (Fig. 3c). This is further demonstrated by histograms of the two maps plotted in Fig. 3d,e. The delocalized CFB has shallow corrugation, leading to a narrower d$I$/d$V$ intensity histogram with a smaller maximum/minimum ratio (Fig. 3d) compared to the more corrugated (i.e., localized) RCB state which has a broader $dI/dV$ intensity histogram with more weight in the intensity troughs (Fig. 3e). Point spectroscopy and $dI/dV$ mapping performed at $V_G = -60$ V allows the VFB peak to be analyzed in the same way since this gate voltage places it in our "elastic window". The VFB state is observed to be similarly delocalized compared to a "remote valence band" (RVB) peak that is localized in the ABBC region analogous to the RCB peak (Supplementary Fig. 2).

To further understand the electronic structure of tDBLG, we methodically tuned the doping level of the system by varying the gate voltage and traced the evolution of $dI/dV$ spectroscopic features. Figure 4d-f shows density plots of $dI/dV$ spectra measured as a function of gate voltage over the range $-60$ V $< V_G < 60$ V for all three stacking regions (ABBC, ABCA, and ABAB). The relationship between the gate voltage $V_G$ and the filling factor $\nu$ (the average number of electrons or holes per moiré unit cell) is found to be $\Delta \nu / \Delta V_G \approx 0.09$ V$^{-1}$ based on the calculated capacitance of the dielectric layers (Methods). The separation of CFB and VFB peaks is clearly observed in the ABAB region for 5 V $< V_G < 35$ V (Fig. 4f). At $V_G = 45$ V ($\nu = 4$) the CFB state is fully filled and the RCB peak appears in the $dI/dV$ spectra, as shown in Fig. 4a-c. Further increasing the gate voltage quickly alters the chemical potential and causes the CFB peak to rapidly drop from the Fermi level as the RCB state begins to be populated with electrons. A similar rapid change occurs at $V_G = -45$ V ($\nu = -4$) as the VFB state becomes fully depleted and the RVB state begins to be populated by holes at lower gate voltages (Fig. 4g-i).

Probably the most significant feature in Fig. 4d-f, however, is that the CFB peak splits into two branches labeled CFB+ and CFB− (with a corresponding dip at the Fermi level) for $V_G \approx 22.5$ V (corresponding to $\nu \approx 2$ and $E \approx 0.12$ V/nm), as highlighted by the dashed black boxes. We observed consistent CFB peak-splitting at $\nu \approx 2$ in several devices, with the local twist angle $\theta$ ranging from 1.05° to 1.17° (see Supplementary Note 2 and Supplementary Fig. 3). The dependence of the splitting on $\nu$ and $E$ confirms that it corresponds to the $\nu = 2$ correlated phase reported in transport measurements[16-20] (see Supplementary Note 3 and Supplementary Fig. 4). The evolution of the CFB with filling factor can be seen even better in Fig. 5a which shows d$I$/d$V$ spectra obtained in the ABBC region over the range $1 < \nu < 3$ (11.5 V < $V_G$ < 33.5 V). A single peak is seen for filling factors away from $\nu = 2$ (e.g. for $V_G$ < 15 V and $V_G$ > 30 V), but a dip feature is clearly resolved at the Fermi level over the range $1.65 \leq \nu \leq 2.35$ (19 V $\leq V_G \leq$ 26 V). By fitting the d$I$/d$V$ signal with the sum of two Lorentzians we are able to extract the magnitude of the energy-splitting $\delta$ as a function of filling factor in the different regions of the moiré unit cell (Fig. 5b; see Supplementary Note 4 and Supplementary Fig. 5 for details). The splitting in the ABBC region reaches its maximum value of $\delta_{max}$ = 18.9 ± 1.2 meV at $\nu \approx 2$ and then decreases to 13 ± 1 meV at $\nu \approx 1.7$ and $\nu \approx$ 2.3 (Fig. 5b blue dots). Beyond this doping range a smaller splitting may still occur, but cannot be determined due to the width of the CFB peak. d$I$/d$V$ spectra measured in the ABCA and ABAB regions display similar trends with $\delta_{max}$(ABCA) = 20.0 ± 1.2 meV and $\delta_{max}$(ABAB) = 20.4 ± 1.3 meV (Fig. 5b purple and orange dots; also see Supplementary Fig. 6). The spatial dependence of the splitting is further illustrated by Fig. 5c which shows d$I$/d$V$ spectra measured at $\nu \approx 2$ along a line cut through the entire moiré unit cell. The CFB+ and CFB− peaks and the dip feature at the Fermi level persist throughout the entire moiré unit cell and do not appear to depend strongly on local stacking order.

Our experimental observations can be understood through comparison to a continuum theoretical model of tDBLG[1,27,31] that includes an added interlayer potential difference and a manually adjusted chemical potential to account for gate-induced variation in $E$-field and carrier density (Methods). Figure 2e shows the band structure calculated at $E = 0$ (corresponding to $V_G = 0$ V) along the high symmetry directions of the moiré Brillouin zone. Two narrow and partially overlapping mini-bands can be seen near the Fermi level (green and pink curves) that are isolated in energy from other bands. Their van Hove singularities result in the CFB and VFB peaks in the calculated LDOS curves of Fig. 2b. For higher values of $E$-field and n-type doping (corresponding to a positively increasing gate voltage) the conduction and valence flat bands both move downward relative to the Fermi level and their energy separation becomes larger (Fig. 2c,d). The corresponding gate-dependent theoretical LDOS curves in Fig. 2b reasonably reproduce the experimental spectra of Fig. 2a, thus establishing the gate-induced $E$-field as the origin of the observed separation of the CFB and VFB peaks.

The spatial delocalization/localization behavior of the CFB/RCB states is also nicely reproduced by the theoretical LDOS calculated at an $E$-field and chemical potential corresponding to the experimental gate voltage $V_G = 60$ V (Fig. 3f-j). Since STS is most sensitive to the electronic states at the sample surface, we plot the calculated LDOS only at the topmost graphene layer to best compare with our measurements (the corresponding behavior when the LDOS interior to the layers is accounted for is discussed in Supplementary Note 5). The spectral density of the CFB peak at $\mathscr{E} = -13$ meV and the RCB peak at $\mathscr{E} = 2.8$ meV in the theoretical LDOS of Fig. 3f correspond well to their experimental counterparts in Fig. 3a (these peaks can be matched to van Hove singularities in the band structure of Fig. 2c). A theoretical LDOS map at the RCB peak energy (Fig. 3h) also agrees with the corresponding experimental d$I$/d$V$ map (Fig. 3c) as the LDOS is seen to concentrate in the ABBC region of the moiré unit cell. This is also reflected in the broad theoretical histogram of the LDOS map (Fig. 3j) which shows reasonable agreement with the experimental histogram (Fig. 3e). The theoretical map of the CFB (Fig. 3g) similarly shows reasonable agreement with the experimental d$I$/d$V$ map of Fig. 3b. Both show strong delocalization across the moiré unit cell as reflected in the narrow histogram of LDOS values seen both in theory (Fig. 3i) and experiment (Fig. 3d). Some discrepancy between experiment (Fig. 3b) and theory (Fig. 3g), however, is seen in the precise locations of minima and maxima for the CFB LDOS, possibly due to approximations made in the continuum model. A similar level of agreement is seen between theory and experiment for the delocalization/localization behavior exhibited by the VFB/RVB electronic features, as shown in Supplementary Fig. 2.

While the model band structure for tDBLG presented above captures the basic experimental electronic structure, it is unable to explain correlation effects such as the splitting observed experimentally at $\nu = 2$. This limitation is evident when we compare the gate-dependent d$I$/d$V$ spectra in Fig. 4d-f to the calculated gate-dependent LDOS in the ABBC, ABCA, and ABAB regions (Supplementary Fig. 8). The theoretical model reproduces many of the experimentally observed features, including the separation of CFB and VFB features for $V_G > 0$ (see Supplementary Note 6 and Supplementary Fig. 9), but the splitting of the CFB peak at $\nu \approx 2$ ($V_G \approx 22.5$ V) is not seen. This is not surprising since the single-particle-level treatment does not include electron-electron interactions and thus cannot capture correlated electronic behavior.

To understand the nature of the correlated state at $\nu \approx 2$ we take electron-electron interactions into account within the self-consistent Hartree-Fock approximation (Methods).[32] We assume that the Coulomb interaction, which is screened by the graphene, the hBN/SiO$_2$ substrate, and the metallic STM tip, takes the single-plane-screened form

$V(\boldsymbol{q}) = \dfrac{e^2}{2\varepsilon_{\text{eff}}\varepsilon_0 q}\Big[1 - \exp\big(-2qd_{\text{S}}\big)\Big]$ where $\varepsilon_{\text{eff}}$ is the effective dielectric constant and $d_{\text{S}}$ is the effective macro tip-sample separation (these are treated as fitting parameters, see Supplementary Note 7). Consistent with the theoretical results of Ref.[27], under these conditions the system can lower its total energy through spontaneous symmetry breaking around $\nu = 2$, with the precise nature of the broken symmetry depending on the detailed band structure and the screening parameters $\varepsilon_{eff}$ and $d_S$. When screening is weak and the electron-electron interaction strength is much greater than the flat band width then the ground state of the system favors isospin polarization (ISP) (it can be either spin-polarized, valley-polarized, or spin-valley-locked, all of which are degenerate in energy within our model). On the other hand, when electron-electron interactions are weak then the development of inter-valley coherence (IVC) is favored.

The different correlated ISP and IVC scenarios for tDBLG lead to very similar results for basic STM observables such as wavefunction maps and energy-dependent LDOS spectra. Nevertheless, we find some quantitative indications for ISP behavior over IVC behavior due to the size of the energy-splitting that we observe experimentally at $\nu = 2$, and so we focus on the tDBLG ISP solution in what follows (ISP versus IVC results are directly compared in Supplementary Note 8). Figure 5d,e shows the ISP Hartree-Fock band structure and density of states (DOS) at $\nu = 2$ for the best fit parameters $\varepsilon_{eff} = 14$ and $d_S = 50$ nm. For these parameters the two isospin sub-bands of the CFB are split by an average of 19 meV across the moiré Brillouin zone, in good agreement with our experimental results (Figure 5a,b). For other bands outside of the CFB and away from the chemical potential the isospin-splitting becomes less significant ($< 5$ meV) and the single-particle band structure is approximately preserved. To better compare the Hartree-Fock calculations with our STS measurements we plot the theoretical isospin-splitting (averaged over the mini-Brillouin zone) as a function of the filling factor (Fig. 5f), and the energy-dependent LDOS at different locations in the moiré unit cell for $\nu = 2$ (Fig. 5g). The largest splitting appears at $\nu = 2$ and the resulting dip feature is observed to persist across the entire moiré unit cell, consistent with the delocalized experimental results shown in Fig. 5b,c. We note, however, that in the Hartree-Fock DOS the splitting of the CFB peak occurs over a wider doping range ($0.5 \leq \nu \leq 3.25$, Supplementary Fig. 11e) than the experimental range $1.65 \leq \nu \leq 2.35$ (Fig. 5a,b). This may be due to the fact that Hartree-Fock is a mean-field theory at zero temperature where quantum and thermal fluctuations are neglected. At finite temperature the splitting is likely to occur over a smaller $\nu$-range due to thermally fluctuating isospin components.

The delocalized spatial distribution of the CFB wavefunctions and the persistent LDOS reduction at $\nu \approx 2$ across the entire moiré unit cell point to a spatially extended correlated state in tDBLG. This result has significant implications for our understanding of electronic correlations in tDBLG and related moiré systems. Metal-insulator transitions in strongly interacting systems often arise from Coulomb repulsion between localized electrons, as exemplified by the single-orbital Hubbard model, and the resulting Mott insulators are usually anti-ferromagnetic due to super-exchange. In the case of tBLG, the localization of low-energy electronic states in the AA stacking region[1,2,21-25] has motivated the application of Hubbard-like models to explain the correlation-driven phases.[33-36] However, recent work has stressed the importance of direct Coulomb exchange and the non-trivial topology of the tBLG bands.[32,35-39] In narrow-band systems with delocalized electron orbitals, direct-exchange can drive ferromagnetic ordering, such as in Stoner ferromagnetism. Theoretical analysis of the tDBLG HF functional[27] indeed shows that a Coulomb scale ($e^2/\varepsilon l_M$) non-local exchange interaction drives symmetry breaking around $\nu = 2$, lifting the degeneracy of the CFB and causing a reduction in the electronic DOS at the Fermi level. Our numerical HF results are found to be consistent with this analysis. This scenario, reminiscent of an integer quantum Hall ferromagnet,[26] is a promising candidate for the observed correlated state in tDBLG.

While our Hartree-Fock calculations qualitatively reproduce the experimental data, they do not resolve the exact nature of the symmetry breaking around $\nu = 2$. For example, the energy difference between the isospin-polarized and the inter-valley coherent states is less than 1 meV per moiré unit cell (Supplementary Fig. 10a), and theoretical LDOS resembling our experimental spectra can also be obtained for the IVC scenario (Supplementary Fig. 12b). In addition, the various states in the isospin-polarized manifold (e.g., spin-polarized, valley-polarized, or spin-valley-locked) are degenerate within our theoretical model and so the specific ground state is likely determined by small effects not accounted for in our model, such as defects and inter-valley Coulomb scattering (see Supplementary Note 8). Transport studies have also resulted in some ambiguity regarding this point, since evidence has been provided that supports both spin-polarized[17-19] and valley-polarized[16] ground states. Nevertheless, in all of these theoretical scenarios the correlated state is the result of non-local exchange interactions and the electron orbitals are spatially-delocalized. Future STM studies involving spin-polarized STM, edge-state detection, and/or quasiparticle interference could potentially provide definitive evidence regarding the symmetry of the ground state and would have significant implications for the pairing channel of any proximate superconductivity.[27]

In conclusion, combined STM/STS measurements reveal a correlated electronic state in tDBLG that is induced by electron-electron interactions when the conduction flat band is separated from the valence flat band and is half-filled. In contrast to tBLG, the flat band wavefunctions in tDBLG are delocalized in space and correlation-driven Fermi-level LDOS reduction is observed everywhere inside the moiré unit cell. A Hartree-Fock analysis of the interacting continuum Hamiltonian shows good agreement with our experimental results and highlights the importance of non-local exchange interactions.

**Methods**

**Sample preparation.** Samples were prepared using the "flip-chip" method[40] followed by a forming-gas anneal.[41,42] Electrical contacts were made by evaporating Cr (5 nm)/Au (50 nm) through a silicon nitride shadow-mask onto the heterostructure. The sample surface cleanliness was confirmed with contact-AFM prior to STM measurements. Samples were annealed at 300 °C overnight in ultra-high vacuum before insertion into the low-temperature STM stage.

**STM/STS measurements.** All STM/STS measurements were performed in a commercial Omicron LT-STM held at $T = 4.7$ K using tungsten (W) tips. STM tips was prepared on a Cu(111) surface and calibrated against the Cu(111) Shockley surface state before each set of measurements to avoid tip artifacts. d$I$/d$V$ spectra were recorded using standard lock-in techniques with a small bias modulation $V_{RMS} = 0.4$–2 mV at 613 Hz. d$I$/d$V$ maps were collected via constant-current grid spectroscopy. All STM images were edited using WSxM software.[43]

**Estimation of carrier density and electric field.** The relation between the gate voltage $V_G$, the carrier density $n$, and the vertical electric field $E$ was estimated by modelling the back-gate configuration in Fig. 1a as a parallel plate capacitor. Therefore

$$n = \frac{\varepsilon_\perp \varepsilon_0 V_G}{e d_D} \tag{1}$$

$$E = \frac{\varepsilon_\perp V_G}{2 d_D} \tag{2}$$

where $d_D = 310$ nm is the thickness of the dielectric layers (hBN and $SiO_2$), $\varepsilon_0$ is the vacuum permittivity, $\varepsilon_\perp \approx 3.5$ is the average perpendicular dielectric constant, and $e$ is the elementary charge. The filling factor $\nu$ and the carrier density $n$ are further related by

$$\nu = \frac{\sqrt{3}}{2} n l_M^2 \tag{3}$$

where $l_M$ is the moiré wavelength. Tip-induced gating effects were not observed in our spectroscopic measurements and thus are not included in the above estimation (see Supplementary Note 9 and Supplementary Fig. 13).

**Continuum model and single-particle calculations.** Our band-structure calculations follow those of Ref.[27], which are based on the Bistritzer-MacDonald continuum approach to moiré structures.[1] The Bernal-stacked bilayer graphene is modeled using a four-band tight-binding model with $t_0 = -2.61$, $t_1 = 0.361$, $t_3 = 0.283$, $t_4 = 0.138$, $\Delta = 0.015$ eV.[44] One bilayer is then rotated by angle $\theta$, and the two bilayers hybridize only through their proximate layers, with intra-sublattice strength $w_0 = 0.075$ eV and inter-sublattice strength $w_1 = 0.1$ eV in the notation of Ref.[27]. We note that in our convention $\theta = 0$ corresponds to the structure obtained when a single sheet of BLG is torn in half, one half translated without rotation, and the two halves stacked with ABBC-alignment. The resulting continuum model is truncated by keeping all states within a radius of 5 mini-Brillouin zones (BZs) of the mini-BZ $\Gamma$-point. The effect of the gate-induced vertical $E$-field is modeled by a constant energy difference $U/3$ between two adjacent layers. To relate $U$ to the applied gate voltage $V_G$ and the physical $E$-field estimated above, we assume an interlayer spacing of $d = 0.33$ nm and model the tDBLG as a uniform dielectric with $\varepsilon_\perp \approx 6.5$, giving $U/eE = 0.15$ nm similar to that reported in Ref.[20].

**Hartree-Fock calculations.** Hartree-Fock calculations are done in a $k$-space approach analogous to earlier Hartree-Fock studies of tBLG[32,38,39] and tDBLG.[27] Our code is an extension of the tBLG code used in Ref.[32]. The Hamiltonian takes the form

$$H = H_0 + PH_\text{C}P - H_\text{BLG}^\text{ex} \qquad (4)$$

where $H_0$ is the continuum band Hamiltonian discussed above, $H_\text{C}$ is the real-space Coulomb interaction $V(q)$, $P$ denotes projection into some number of moiré bands near charge neutrality, and $H_\text{BLG}^\text{ex}$ is a single-particle correction to be discussed shortly. The Coulomb matrix elements for $H_\text{C}$ are evaluated in the basis of the continuum band structure and projected into the 6 bands nearest to the charge neutrality point per valley and spin, for a total of $4 \times 6 = 24$ bands. When evaluating the Coulomb integrals, we ignore the interlayer spacing $d$, which is accurate up to corrections of order $d/l_M$. We note that while small, the neglected $d/l_M$ terms will lead to interlayer screening which modifies the effective $U$, which may be an interesting direction for future work. Since the BLG tight-binding parameters obtained from DFT already contain the effect of renormalization by the Coulomb interaction on the physics of a single bilayer, we follow Refs.[32,38] by subtracting off the Hartree-Fock Hamiltonian $H_{BLG}^{ex}$ (exchange energy) of two decoupled BLG layers at charge neutrality.

We then consider a Slater-determinant ansatz which is diagonal in the mBZ momentum $k$,

$$|u\rangle = \prod_{k \in mBZ} \prod_{j=1}^{n_k} \left( \sum_n u_{k,j}^n c_{k,n}^\dagger \right) |0\rangle \qquad (5)$$

Here $c_{k,n}^\dagger$ creates an electron in eigenstate $(k, n)$ of the band structure, while $u_{k,j}^n$'s are the set of variational parameters to be optimized. The total occupation $n_k$ is allowed to vary across the mBZ to account for the presence of Fermi surfaces. In addition to translation invariance, we constrain the $u$'s to preserve a spin-symmetry about the $S^z$ axis (ruling out non-colinear magnetism, which is not expected in this model), but do not enforce the valley-U(1) symmetry, allowing for spontaneous inter-valley coherence as has been argued to occur in tBLG.[32] Discretizing the model on a $20 \times 20$ $k$-grid, the $u$'s are iteratively adjusted to minimize the energy $\langle u|H|u\rangle$, using the optimal damping algorithm to achieve Hartree-Fock self-consistency. Solving for the self-consistent Hartree-Fock Hamiltonian $H_{HF}(\nu, U)$ at each filling $\nu$ and electric field $U$, we then reconstruct DOS and LDOS curves by diagonalizing $H_{HF}$ and converting back to real space using the continuum-model wavefunctions. To account for finite temperature and instrumental broadening effects and avoid spurious spikes due to the numerical discretization of the mini-BZ, the (L)DOS is broadened by convolving with

$$f(\mathcal{E}, \mathcal{E}') = \frac{2}{\pi\lambda} \left[ \cosh \frac{2(\mathcal{E} - \mathcal{E}')}{\lambda} \right]^{-1}$$ , where $\lambda = 2$ meV for DOS and 4 meV for LDOS. Unless

specified, the LDOS is always projected onto the topmost graphene layer to enable comparison with d$I$/d$V$ spectroscopy.

## Acknowledgements

The authors thank D. Goldhaber-Gordon, J. Jung, A. Vishwanath, D. Wong, A. Sharpe, N. Bultinck, J. Y. Lee, E. Khalaf, H.-Z. Tsai, F. Liou, Y. Chen and G. Wang for helpful conversations. This research was supported by the sp2 program (KC2207) (STM measurement and instrumentation) funded by the Director, Office of Science, Office of Basic Energy Sciences, Materials Sciences and Engineering Division, of the U.S. Department of Energy under Contract No. DE-AC02-05CH11231. For graphene characterization we used the Molecular Foundry at LBNL, which is funded by the Director, Office of Science, Office of Basic Energy Sciences, Scientific User Facilities Division, of the US Department of Energy under Contract No. DE-AC02-05CH11231. Support was also provided by National Science Foundation Award DMR-1807233 (device fabrication, image analysis). K.W. and T.T. acknowledge support from the Elemental Strategy Initiative conducted by the MEXT, Japan, Grant Number JPMXP0112101001, JSPS KAKENHI Grant Number JP20H00354 and the CREST(JPMJCR15F3), JST. S.L. acknowledges support from Kavli ENSI Heising Simons Junior Fellowship. C.H. acknowledges the support of Alexander von Humboldt Foundation for a Feodor Lynen research fellowship.

## Author contributions

## Competing Interests

The authors declare no competing interests.

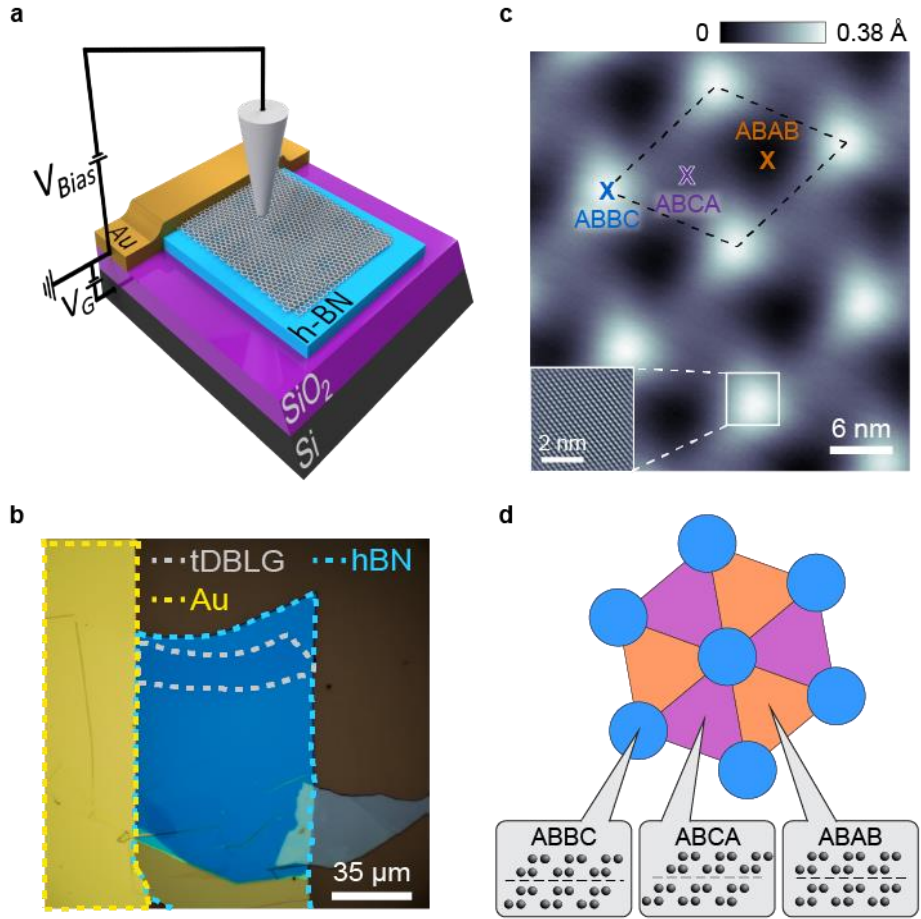

**Figure 1: Basic characterization of tDBLG. a**, STM/STS measurement configuration for tDBLG devices. The doped Si substrate acts as a back-gate to allow tuning of the device carrier density. **b**, Optical microscope image of a tDBLG device. **c**, STM topographic image of tDBLG with a twist angle $\theta \approx 1.08°$ ($V_{\text{Bias}} = -250$ mV, $I_0 = 0.25$ nA). The dashed box outlines the moiré unit cell. Inset: zoom-in image showing graphene atomic lattice. **d**, Sketch of tDBLG moiré pattern showing three distinct stacking orders. The inner two layers of carbon atoms are on top of each other for ABBC stacking whereas they are Bernal-stacked for ABCA and ABAB stacking.

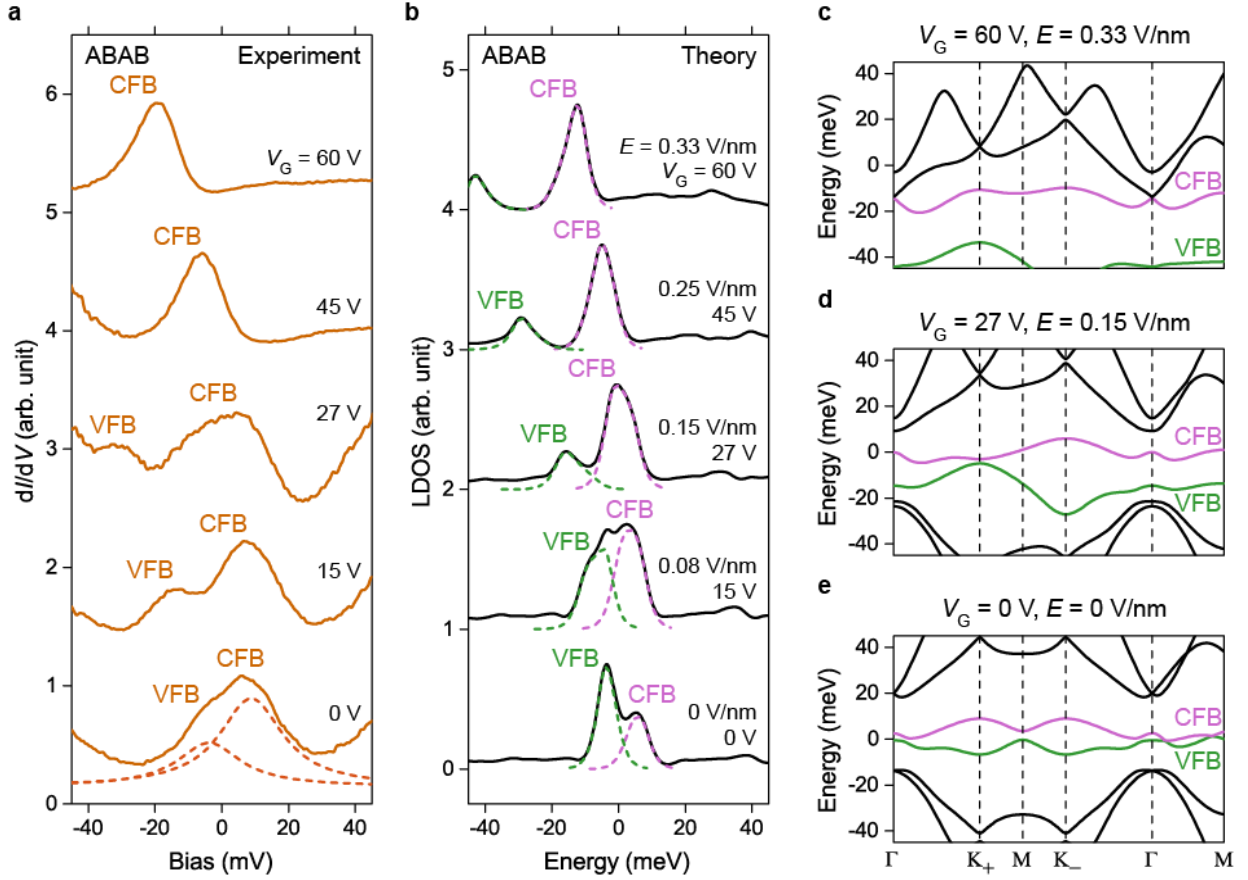

**Figure 2: Electric field tuning of tDBLG electronic structure**. **a**, d$I$/d$V$ spectra measured in an ABAB region under different back-gate voltages (modulation voltage $V_{RMS}$ = 1 mV; initial $V_{Bias}$ = −100 mV, $I_0$ = 0.5 nA). **b**, Theoretical LDOS for ABAB region for gate-voltages in (**a**) (corresponding $E$-fields are shown). The dashed curves denote contributions from each separate flat band. **c-e**, Single-particle band structure along the high symmetry directions of the tDBLG moiré Brillouin zone for (**c**) $V_G$ = 0, $E$ = 0; (**d**) $V_G$ = 27 V, $E$ = 0.15 V/nm; and (**e**) $V_G$ = 60 V, $E$ = 0.33 V/nm. For clarity only bands from a single valley are shown. CFB = conduction flat band; VFB = valence flat band.

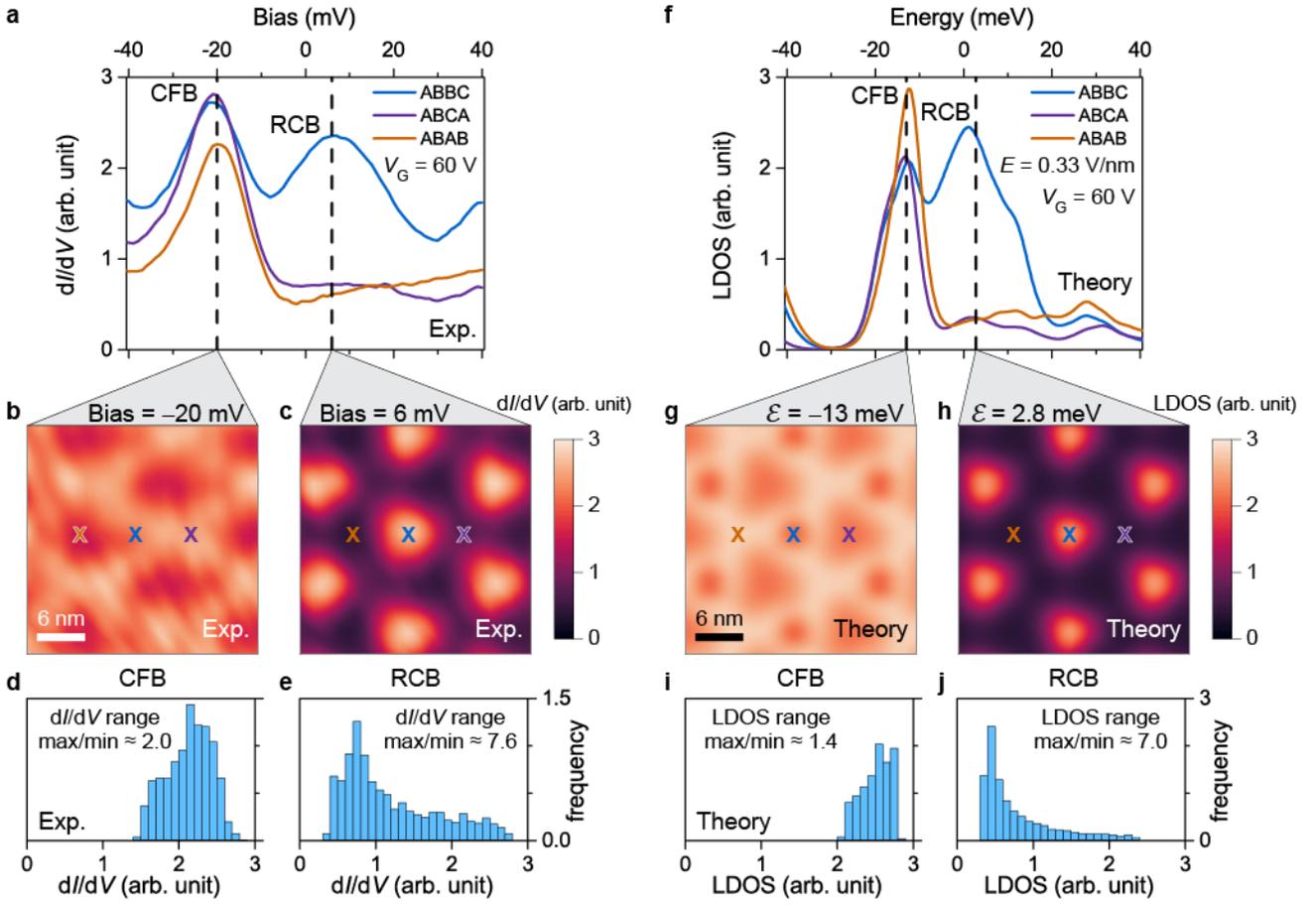

**Figure 3: Spatial distribution of conduction flat band and remote conduction band wavefunctions. a**, $dI/dV$ spectra for three different stacking regions measured at $V_G = 60$ V. **b, c**, $dI/dV$ maps of the same tDBLG region obtained at (**b**) $V_{Bias} = -20$ mV and (**c**) $V_{Bias} = 6$ mV (modulation voltage $V_{RMS} = 1$ mV; initial $V_{Bias} = -500$ mV, $I_0 = 2.5$ nA). **d**, Histogram of (**b**). **e**, Histogram of (**c**). **f**, Theoretical LDOS of three different stacking regions for $V_G = 60$ V and $E = 0.33$ V/nm using a single-particle continuum model. **g, h**, LDOS maps at energies (**g**) –13 meV and (**h**) 2.8 meV. **i**, Histogram of (**g**). **j**, Histogram of (**h**). All histograms use bin size of 0.1 arb. unit. Areas under histograms are normalized to 1. CFB = conduction flat band; RCB = remote conduction band.

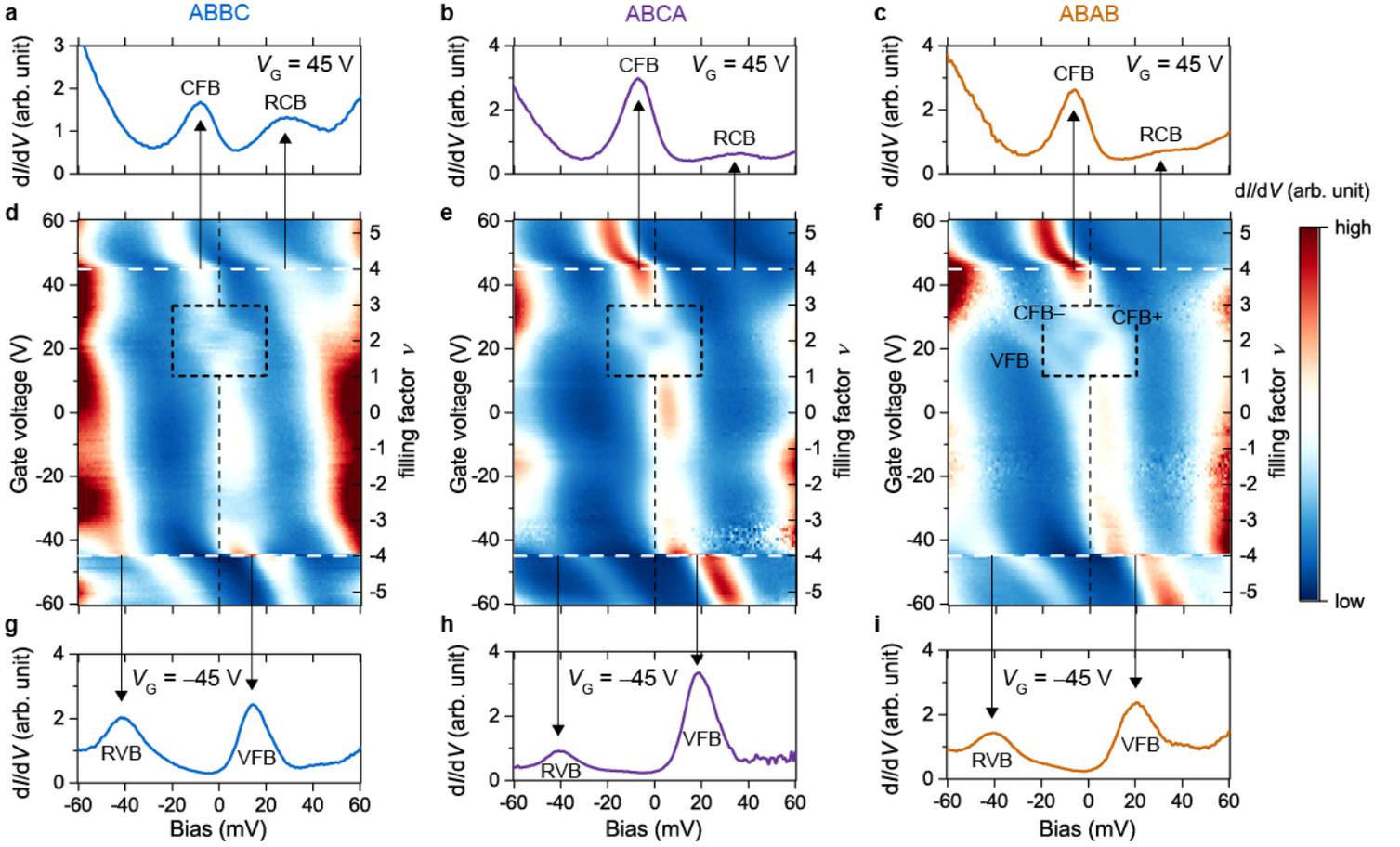

**Figure 4: Gate-dependent d$I$/d$V$ spectroscopy for three different stacking regions. a-c**, d$I$/d$V$ spectra for three stacking regions for $V_G$ = 45 V ($\nu$ = 4). **d-f**, Gate-dependent d$I$/d$V$ density plot for three stacking regions over gate-voltage range –60 V < $V_G$ < 60 V. The vertical black dashed line denotes zero bias (the Fermi level). The black dashed box highlights correlation-driven splitting of the CFB peak near $\nu \approx 2$. **g-i**, d$I$/d$V$ spectra for three stacking regions for $V_G$ = –45 V ($\nu$ = –4). Spectroscopy parameters: modulation voltage $V_{RMS}$ = 1 mV; initial $V_{Bias}$ = –100 mV, $I_0$ = 0.5 nA. CFB = conduction flat band; VFB = valence flat band; RCB = remote conduction band; RVB = remote valence band.

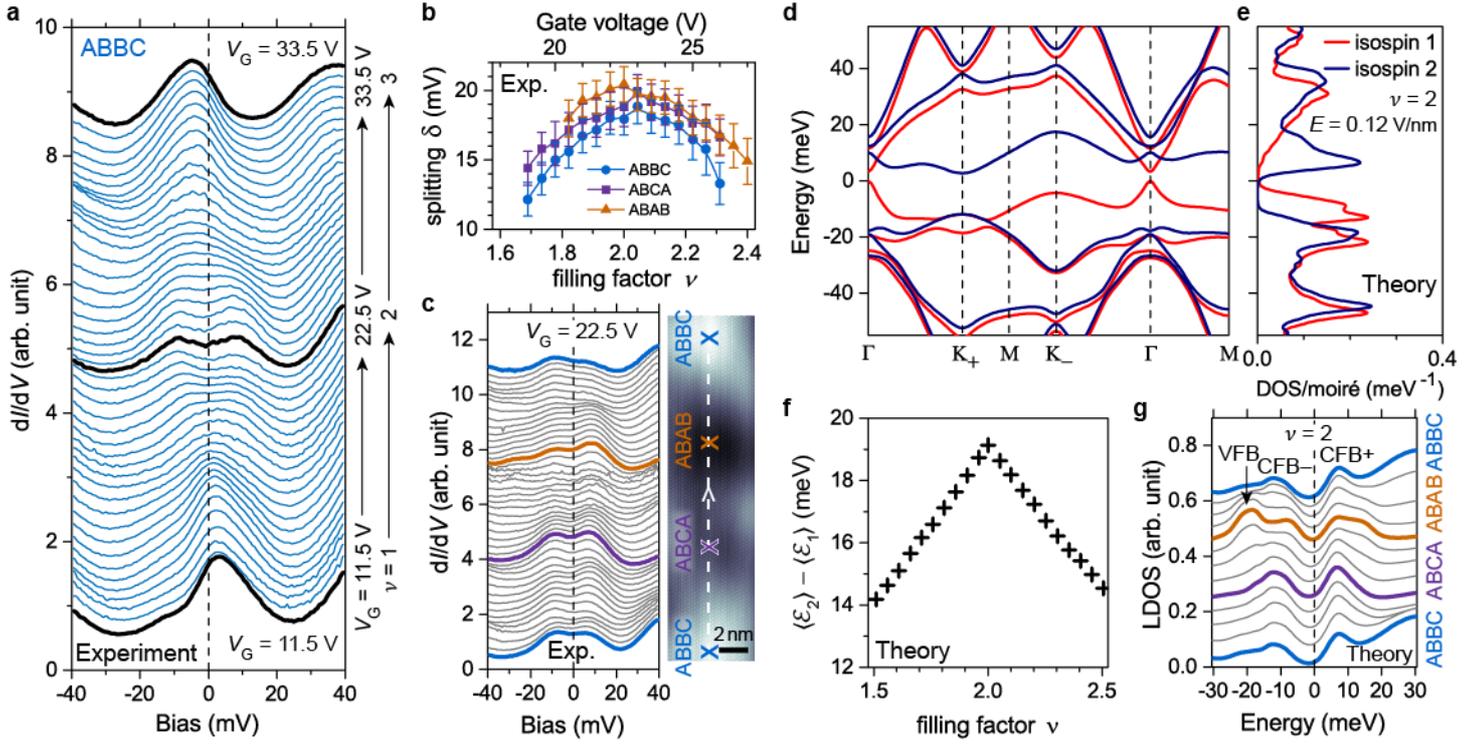

**Figure 5: Emergent correlated state at $\nu \approx 2$. a,** Gate-dependent d$I$/d$V$ spectra in ABBC region for 11.5 V < $V_\text{G}$ < 33.5 V (1 < $\nu$ < 3) (modulation voltage $V_\text{RMS}$ = 1 mV; initial $V_\text{Bias}$ = –100 mV, $I_0$ = 0.5 nA). **b,** Magnitude of CFB energy-splitting extracted as a function of gate voltage (filling factor) for three stacking regions. The error bars were estimated by combining fitting uncertainty, finite temperature broadening, and an instrumental broadening of ~1 mV. **c,** Spatially-resolved d$I$/d$V$ spectra measured at $V_\text{G}$ = 22.5 V ($\nu$ = 2) along the white dashed line in the topographic image (which goes through all three tDBLG stacking regions). Scanning parameters: $V_\text{Bias}$ = –100 mV, $I_0$ = 0.1 nA. Spectroscopy parameters: modulation voltage $V_\text{RMS}$ = 1 mV; initial $V_\text{Bias}$ = –100 mV, $I_0$ = 0.5 nA. **d,** Hartree-Fock band structure of tDBLG and **e,** the corresponding isospin-resolved DOS for an isospin-polarized solution at $\nu$ = 2 ($E$ = 0.12 V/nm, $\varepsilon_\text{eff}$ = 14, $d_\text{S}$ = 50 nm). For clarity only bands from a single valley are shown. **f,** Energy offset (i.e., splitting) between the isospin sub-bands averaged over the mini-Brillouin zone as a function of filling factor for 1.5 < $\nu$ < 2.5. **g,** Calculated Hartree-Fock LDOS for $\nu$ = 2 at different positions across the moiré unit cell corresponding to white dashed line in (**c**). CFB = conduction flat band; VFB = valence flat band.